\documentclass[sigconf, language=french,
language=german, language=spanish, language=english]{acmart}

\AtBeginDocument{%
  \providecommand\BibTeX{{%
    \normalfont B\kern-0.5em{\scshape i\kern-0.25em b}\kern-0.8em\TeX}}}

\setcopyright{acmcopyright}
\copyrightyear{2018}
\acmYear{2018}
\acmDOI{XXXXXXX.XXXXXXX}

\acmConference[Conference’17]{}{July 2017}{Washington, DC, USA}
\acmPrice{15.00}
\acmISBN{978-1-4503-XXXX-X/18/06}

\begin{document}

\title{Beyond Screens: Supporting Co-located Augmented Reality
Experiences with Smart Home Devices}


\author{Ava Robinson, Yu Jiang Tham, Rajan Vaish, Andrés Monroy-Hernández}
\affiliation{%
  \institution{Snap Inc.}
  \country{USA}}
\email{arobinson,yujiang,rvaish,amh@snap.com}

\renewcommand{\shortauthors}{Trovato and Tobin, et al.}

\begin{abstract}
We introduce Spooky Spirits, an AR game that makes novel use of
everyday smart home devices to support co-located play. Recent
exploration of co-located AR experiences consists mainly of digital
visual augmentations on mobile or head-mounted screens. In this
work, we leverage widely adopted smart lightbulbs to expand AR
capabilities beyond the digital and into the physical world, further
leveraging the physicality of users’ shared environment.

\end{abstract}





\maketitle

\section{Introduction}
Recent research has shown how co-located AR experiences can
enable people to have fun together, but most consist of only digital
visual augmentations using phone-based AR \cite{dagan2022project}. Additionally, prior
work has shown that AR is well-suited for co-located experiences
because it can be grounded in the environment \cite{wetzel2008guidelines}, for example
by using physical objects as enablers, or inputs, of the experience
\cite{dagan2022project}. However, these augmentations are constrained to pixels on a
screen rather than augmenting the physical space itself or using
physical objects as outputs in the experience. The recent increase
of IoT devices in people’s homes \cite{koskela2004evolution} presents an opportunity to use
smart home devices to extend the augmentation of the physical
world for co-located experiences.

\begin{figure}[h!] \includegraphics[width=0.48\textwidth]{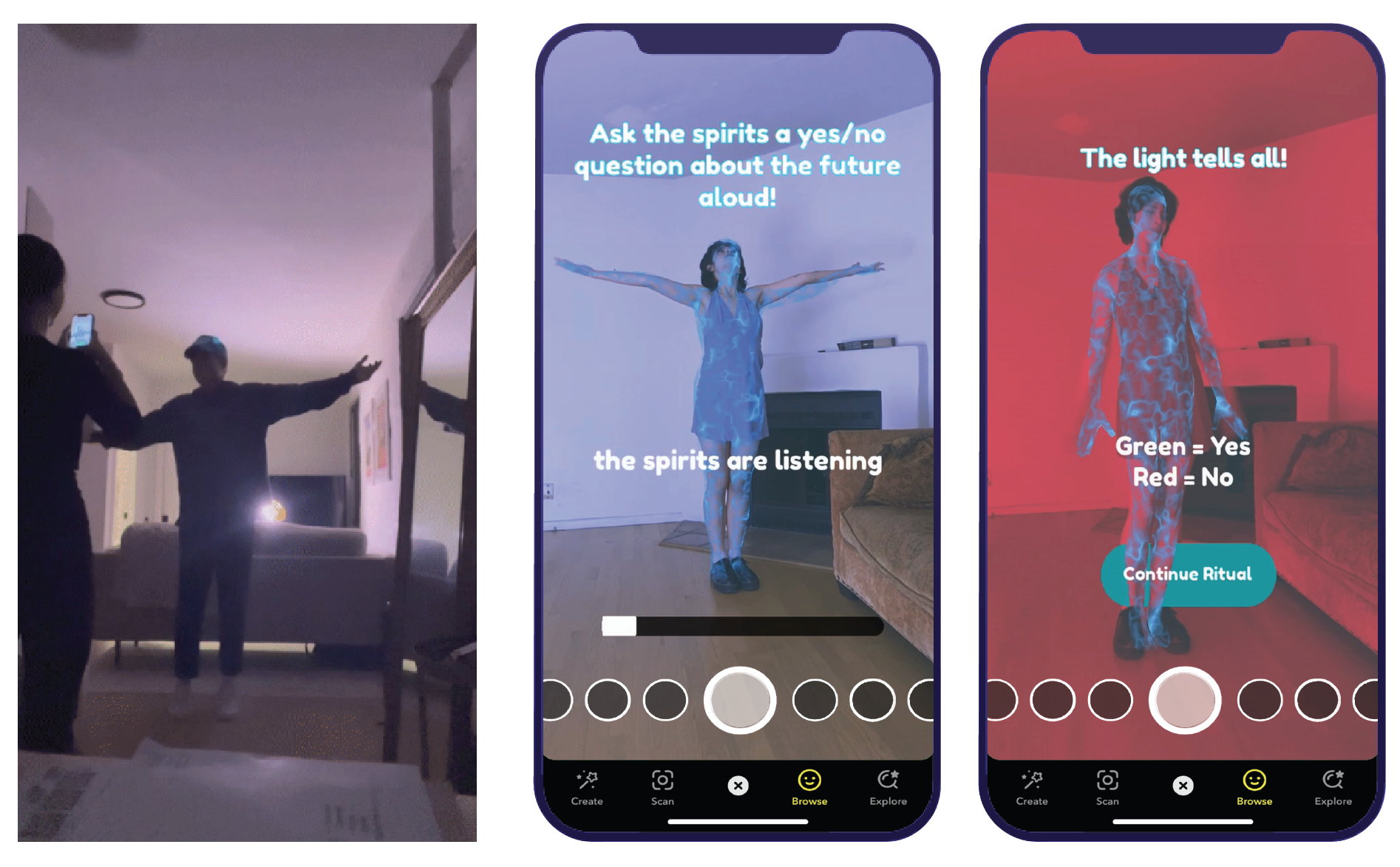}
  \caption{Left: Users playing Spooky Spirits with the smart
light in the background and a user doing the T Pose body
gesture. Middle: Game prompting users to ask a question
while the light illuminates the room white indicating “spirits”
listening. Right: Room getting illuminated red by the light,
representing a “no” answer.
}
  \label{fig1}
\end{figure}

Leveraging this opportunity, we explore using IoT devices, in
particular a smart light, as a core aspect of an experience and novel form of interaction. We present Spooky Spirits, a mobile AR game\footnote{Implemented as a Snapchat Lens using Lens Studio \url{https://ar.snap.com/}} built for two users to play together in person along with a smart
light \footnote{Used Kasa Smart Lightbulbs \url{https://www.kasasmart.com/us/products/smart-lighting}}. Our system allows users to easily connect their smart lights
to the mobile experience via a web interface.

We evaluated our system with 10 pairs of participants by observing how they use the experience and interviewing them to
learn more about the immersive value and impact of augmenting a
shared physical space using an IoT device. We hope to motivate and
inform future designs of mobile AR and IoT co-located experiences.

\section{DESIGN AND EXPERIENCE OVERVIEW}
The Spooky Spirits experience piggybacks off of existing spirit
summoning and fortune-telling games such as Ouija boards\footnote{Ouija board game:\url{https://en.wikipedia.org/wiki/Ouija}}and
Magic 8-Balls\footnote{Magic 8-Ball game: \url{https://en.wikipedia.org/wiki/Magic_8_Ball}}
. This narrative embraces the magic and spookiness
of a smart light changing colors without manual input. To set up
the experience users must 1) have a smart light in their space 2)
connect their light to the mobile game using a website, and 3) find
a partner to play with in person.

Device arrangement is a key design consideration for co-located
experiences \cite{isbister2018social}\cite{lundgren2015designing}. Our experience involves two users, a phone
running the game, and a smart light. When playing Spooky Spirits, one user, player 1, holds the phone with the game running and
guides their partner, player 2, through the playful “summoning
ritual” by communicating the instructions they receive via the
mobile UI. When the experience first begins, the game creates the
spooky ambiance by turning the smart light blue. Player 1 then
directs their partner to stand in front of the camera and position
their body in a specific gesture, for example, a T Pose, which is
detected using Full Body Triggers \footnote{https://docs.snap.com/lens-studio/references/templates/object/full-body-triggers}
(see Fig. 1 - left). After the body
gesture is detected, the light turns white providing visual feedback
for both users indicating that the “spirits” are listening and users are
prompted to ask a yes-no question about the future aloud (see Fig.
1 - Middle). For example, users might ask "Will I win the lottery?".

To make the light a salient part of the experience we chose to
use it not only as an enhancement to ambiance but also as a source
of information in the narrative of the experience. After users ask a
yes-no question aloud, the light will turn green or red representing
the answer as yes or no respectively (see Fig. 1 - Right), making the
experience dependent on the information from the state of the light.
In this design, the smart light acts as the sole source of immersion
and augmentation for player 2. This allows us to explore if the
experience is immersive even without mobile visual augmentation
for all players.

This design encourages users to work together and leverages
the physicality of users’ space by augmenting their environment
beyond pixels on a screen. This provides an opportunity to observe
the impact and immersive value that the smart home device brings
to the experience since only one user can see the UI and AR on the
mobile device, while the other user can only see the effects of the
light in their environment. This experience extends augmentation
beyond mobile for both users.

\section{ SMART DEVICE INTEGRATION SYSTEM}

\begin{figure}[h!] \includegraphics[width=0.3\textwidth]{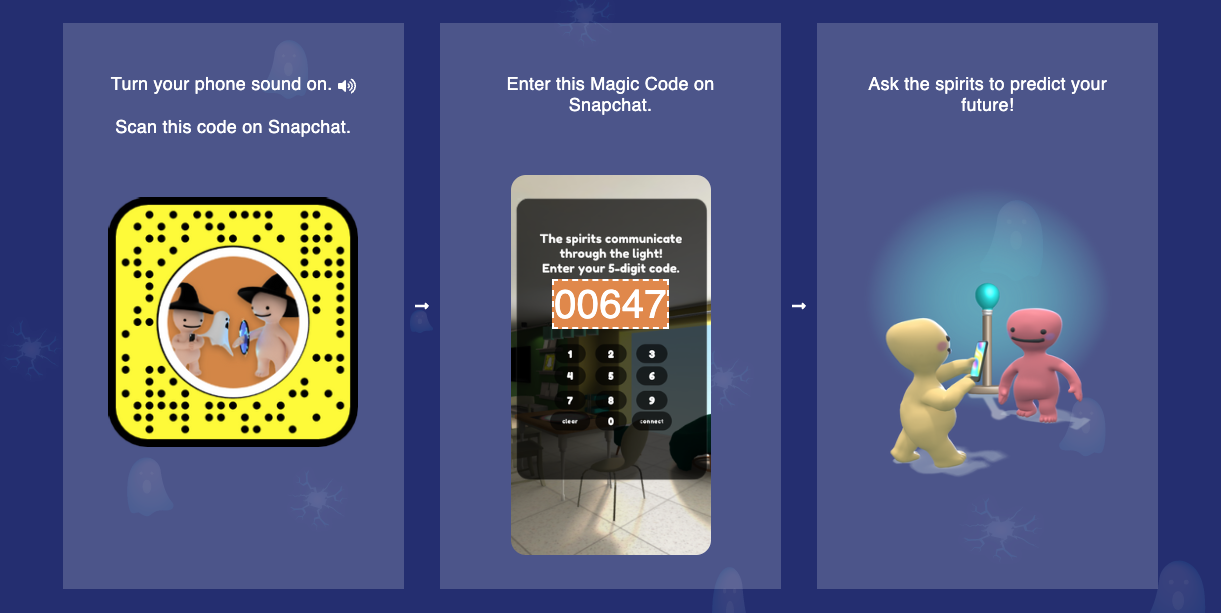}
  \caption{Website for users to launch the game and generate
a 5-digit code to connect to their smart light.}
  \label{fig2}
\end{figure}

\begin{figure}[h!] \includegraphics[width=0.3\textwidth]{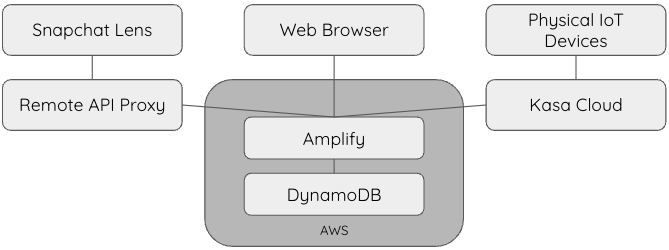}
  \caption{Overview of the backend system for integrating
Kasa Devices into the game.}
  \label{fig3}
\end{figure}

Our system involves a web app\footnote{https://spookyspirits.letsplayirl.com/} we built for users to connect to
their smart lights (see Fig. 2). We chose to use Kasa Smart Light
Bulbs as they have a public API, are easy for non-technical users to
set up, and are relatively inexpensive. As set up for our experience,
users use our web interface to log in to their existing Kasa account,
select the bulb they would like to use, and obtain a 5-digit randomly
generated code that is stored in DynamoDB. Next, users launch
the game\footnote{They launched by scanning a code in Snapchat}
and enter the 5-digit code to connect the game to their
smart light. By using an external interface for login we hoped
to simplify the connection process and reduce friction within the
mobile experience. The details of the system consist of AWS Amplify
running a Next.js app with 1) a server-side component that handles
API calls to get and set the settings of the smart light and 2) a client side component that displays the website to the user (see Fig. 3).
The Next.js app also polls from a DynamoDB database periodically
to check for changes for each device, represented by the generated
5-digit code, and if there are changes, it sends the new data to the
Kasa Cloud using the Kasa public API, which modifies the state of
the physical devices. The Snapchat Lens will make get and set calls
to the AWS Remote API using the entered 5-digit code to change
the state of the device.

\section{USER STUDY OBSERVATIONS}

We performed a pilot study with 10 pairs of participants playing
with the Spooky Spirits experience together with a smart light and
conducted semi-structured interviews.

We aimed to explore the impact of a smart home device as a core
aspect of a co-located AR experience to help inform future designs
of similar experiences. Here are six main observations

\begin{enumerate}
\item  Augmentation of the physical world beyond mobile superseded the setup effort. Users noted that the setup friction
was high due to the smart light, however, they felt that the
light added to the overall immersive value.
\item  Augmentation via IoT devices enabled immersion beyond mobile and overcame asymmetric device access. Although one
user did not experience mobile visual augmentation, generally they still felt immersed and involved in the experience
because of the augmentations from the smart light.
\item  Different setup decisions led to new interactions. We observed
users who placed the bulb in a ceiling socket try to reach up
towards the light and pay direct attention to the bulb itself,
whereas users who placed the bulb in a desk or standing
lamp did not try to touch the physical bulb and instead paid
attention to the effect the bulb had on overall lighting.
\item  The state of the physical environment affected its augmentation. Users who played in dark lit rooms felt the experience
was more immersive than those who played in rooms with
more secondary light because the changes in the color and
brightness of the light were more salient.
\item  IoT devices could have played a more important role in giving feedback. Users sometimes struggled to know what to
expect from the smart light or when to wait for the light to
change.

\item  IoT devices were able to serve multiple roles, from creating
ambiance to providing information. During the game, blue
light was used to create spooky ambiance, and green or
red light encoded information (yes or no answers to users’
questions) that users easily decoded.
\end{enumerate}
In the future, we hope to build and study more co-located AR
experiences that use a variety of IoT devices as novel forms of
augmentation and interactions. We hope this inspires future designers to continue to design more mobile AR and IoT co-located
experiences.

\begin{acks}
Many thanks to Melissa Powers, Erica Principe Cruz, Samantha
Reig, Tim Chong, Jennifer He, and Eunice Kim for their help and support in designing, building, and studying this experience. Also,
many thanks to all our study participants for their time.

\end{acks}

\bibliographystyle{ACM-Reference-Format}
\bibliography{final}


\begin{thebibliography}{5}


\ifx \showCODEN    \undefined \def \showCODEN     #1{\unskip}     \fi
\ifx \showDOI      \undefined \def \showDOI       #1{#1}\fi
\ifx \showISBNx    \undefined \def \showISBNx     #1{\unskip}     \fi
\ifx \showISBNxiii \undefined \def \showISBNxiii  #1{\unskip}     \fi
\ifx \showISSN     \undefined \def \showISSN      #1{\unskip}     \fi
\ifx \showLCCN     \undefined \def \showLCCN      #1{\unskip}     \fi
\ifx \shownote     \undefined \def \shownote      #1{#1}          \fi
\ifx \showarticletitle \undefined \def \showarticletitle #1{#1}   \fi
\ifx \showURL      \undefined \def \showURL       {\relax}        \fi
\providecommand\bibfield[2]{#2}
\providecommand\bibinfo[2]{#2}
\providecommand\natexlab[1]{#1}
\providecommand\showeprint[2][]{arXiv:#2}

\bibitem[Dagan et~al\mbox{.}(2022)]%
        {dagan2022project}
\bibfield{author}{\bibinfo{person}{Ella Dagan}, \bibinfo{person}{Ana~Mar{\'\i}a
  C{\'a}rdenas~Gasca}, \bibinfo{person}{Ava Robinson}, \bibinfo{person}{Anwar
  Noriega}, \bibinfo{person}{Yu~Jiang Tham}, \bibinfo{person}{Rajan Vaish},
  {and} \bibinfo{person}{Andr{\'e}s Monroy-Hern{\'a}ndez}.}
  \bibinfo{year}{2022}\natexlab{}.
\newblock \showarticletitle{Project IRL: Playful Co-Located Interactions with
  Mobile Augmented Reality}.
\newblock \bibinfo{journal}{\emph{Proceedings of the ACM on Human-Computer
  Interaction}} \bibinfo{volume}{6}, \bibinfo{number}{CSCW1}
  (\bibinfo{year}{2022}), \bibinfo{pages}{1--27}.
\newblock


\bibitem[Isbister et~al\mbox{.}(2018)]%
        {isbister2018social}
\bibfield{author}{\bibinfo{person}{Katherine Isbister}, \bibinfo{person}{Elena
  M{\'a}rquez~Segura}, {and} \bibinfo{person}{Edward~F Melcer}.}
  \bibinfo{year}{2018}\natexlab{}.
\newblock \showarticletitle{Social affordances at play: Game design toward
  socio-technical innovation}. In \bibinfo{booktitle}{\emph{Proceedings of the
  2018 CHI Conference on Human Factors in Computing Systems}}.
  \bibinfo{pages}{1--10}.
\newblock


\bibitem[Koskela and V{\"a}{\"a}n{\"a}nen-Vainio-Mattila(2004)]%
        {koskela2004evolution}
\bibfield{author}{\bibinfo{person}{Tiiu Koskela} {and} \bibinfo{person}{Kaisa
  V{\"a}{\"a}n{\"a}nen-Vainio-Mattila}.} \bibinfo{year}{2004}\natexlab{}.
\newblock \showarticletitle{Evolution towards smart home environments:
  empirical evaluation of three user interfaces}.
\newblock \bibinfo{journal}{\emph{Personal and Ubiquitous Computing}}
  \bibinfo{volume}{8} (\bibinfo{year}{2004}), \bibinfo{pages}{234--240}.
\newblock


\bibitem[Lundgren et~al\mbox{.}(2015)]%
        {lundgren2015designing}
\bibfield{author}{\bibinfo{person}{Sus Lundgren}, \bibinfo{person}{Joel~E
  Fischer}, \bibinfo{person}{Stuart Reeves}, {and} \bibinfo{person}{Olof
  Torgersson}.} \bibinfo{year}{2015}\natexlab{}.
\newblock \showarticletitle{Designing mobile experiences for collocated
  interaction}. In \bibinfo{booktitle}{\emph{Proceedings of the 18th ACM
  conference on computer supported cooperative work \& social computing}}.
  \bibinfo{pages}{496--507}.
\newblock


\bibitem[Wetzel et~al\mbox{.}(2008)]%
        {wetzel2008guidelines}
\bibfield{author}{\bibinfo{person}{Richard Wetzel}, \bibinfo{person}{Rod
  McCall}, \bibinfo{person}{Anne-Kathrin Braun}, {and}
  \bibinfo{person}{Wolfgang Broll}.} \bibinfo{year}{2008}\natexlab{}.
\newblock \showarticletitle{Guidelines for designing augmented reality games}.
  In \bibinfo{booktitle}{\emph{Proceedings of the 2008 Conference on Future
  Play: Research, Play, Share}}. \bibinfo{pages}{173--180}.
\newblock


\end{thebibliography}

\end{document}